\documentstyle[aps,preprint,amssymb,eqsecnum,tighten]{revtex}
\begin{document}
\draft

\title{Black Hole Data via a Kerr-Schild Approach}

\author{Nigel T. Bishop${}^{1}$, Richard Isaacson${}^{2,3}$,
Manoj Maharaj${}^{4}$ and Jeffrey Winicour${}^{2}$}

\address{${}^{1}$Department of Mathematics, Applied Mathematics and Astronomy,
University of South Africa, P.O. Box 392, Pretoria 0003, South Africa \\
${}^{2}$Department of Physics and Astronomy,
University of Pittsburgh, Pittsburgh, PA 15260\\
${}^{3}$Physics Division, National Science Foundation,
Arlington VA22230\\
${}^{4}$Department of Mathematics and Applied Mathematics,
University of Durban-Westville, Durban 4000, South Africa}

\maketitle

\begin{abstract}
We present a new approach for setting initial Cauchy data for multiple black
hole spacetimes. The method is based upon adopting an initially Kerr-Schild
form
of the metric. In the case of non-spinning holes, the constraint equations
take a simple hierarchical form which is amenable to direct numerical
integration. The feasibility of this approach is demonstrated by
solving analytically the problem of initial data in a perturbed Schwarzschild
geometry.
\end{abstract}

\pacs{04.20Ex, 04.25Dm, 04.25Nx, 04.70Bw}

\section{Introduction}

The calculation of gravitational radiation from the inspiral of binary
black holes is of paramount importance to the development of
gravitational wave detection into a new form of astronomy. The strong
field regime of the emission process occurring during the merger of two
black holes can only be treated by numerical evolution. However, in the
near future it is unclear whether computing power and techniques will be
adequate to track the two black holes through many orbits, beginning
with a quasi-Newtonian orbit which can be approximated by perturbation
methods and ending with their merger into a single black hole. Instead,
the merger radiation may have to be explored beginning with initial
data in a non-perturbative setting just prior to the final orbital
plunge. Numerical codes are presently being developed which promise to
provide an accurate map between such initial data and the emitted
gravitational waveform. However, the extent to which the merger
waveform will have a characteristic signature independent of the
details of the initial data is unknown, and there is some evidence that
the assumptions that have to be made in the usual approach lead to data
that is not physically realistic~\cite{lous}. For this reason it is
imperative that many different forms of black hole initial data be
available.

Results on multiple black hole initial data were presented by Misner in
1963~\cite{mis}. As is well known he found an analytic solution for the
special case of a hypersurface of time-symmetry. In this case the
extrinsic curvature is zero so the momentum constraints are satisfied
identically; and, assuming conformal flatness of the initial 3-geometry,
the Hamiltonian constraint reduces to Laplaces's
equation for the conformal factor. Subsequent work has been essentially
a generalization of Misner's result. The first step in treating the
non-time-symmetric case was the development of the theory of conformal
decomposition~\cite{york}. This then led to methods for determining
initial data for one black hole with specified linear and angular
momentum~\cite{bow}.  A procedure for ``summing'' such solutions has
been described~\cite{kul}, and the procedure has been
implemented numerically~\cite{cook1,cook2}. Recently, a modified version
of this procedure has been formulated which avoids complications at the
inner boundary for this elliptic problem by compactifying the region
interior to the black holes~\cite{bb}. However, only conformally flat
geometries with trace-free extrinsic curvatures can be readily obtained
by this technique.

Here we describe the generation of initial Cauchy data by means of a
new approach, based upon adoption of an instantaneously
Kerr-Schild~\cite{ks1,ks2}
form of the metric on the initial time slice. In this approach the
time-symmetric Misner data is not expressible as a simple special case.
For a single non-spinning vacuum black hole at rest, it leads to
Schwarzschild initial data in Eddington-Finklestein coordinates. In the
numerical treatment of the non-vacuum spherically symmetric
gravitational collapse of a scalar field, the most stable Cauchy
evolution algorithms for the inner boundary at the apparent horizon
have been based upon a generalization of ingoing Eddington-Finklestein
coordinates~\cite{mc}. An analogous gauge is likely to be beneficial to
stability
in the two black hole case. Since Eddington-Finklestein coordinates are
closely related to Kerr-Schild coordinates, the initial data generated
by our approach offers the additional advantage of allowing greater
flexibility in the choice of gauge conditions on the lapse and shift in
the search for stable numerical algorithms for the simulation of black
holes. Similar motivations have led to another independent formulation
of the initial data problem which reduces to the Kerr-Schild form for
the Schwarzschild or Kerr cases~\cite{matzn}.

The framework used is that of an instantaneously Kerr-Schild
space-time, which is defined and discussed in section~\ref{sec:inst}.
The corresponding initial data is more general than for exact Kerr-Schild
space-times, and the constraint equations are derived from first
principles in section~\ref{sec:cons}. A numerical procedure for solving
the constraint equations is described in section~\ref{sec:sol}; also,
we show that a simple algebraic condition on the metric defines the
location of the apparent horizon. We do not, in this paper, solve the
numerical problem; but we show in section~\ref{sec:pert} that our
solution procedure is feasible by formulating, and then solving
analytically, the Schwarzschild perturbation problem corresponding to
the ``close approximation'' for two black holes~\cite{pp}. We conclude
in section~\ref{sec:conc} with comments on the future development of
this research.

\section{Instantaneously Kerr-Schild space-times}
\label{sec:inst}

In the 1960s the work of Trautman~\cite{trau}, and of Kerr and
Schild~\cite{ks1,ks2}, led to the development of the theory of Kerr-Schild
geometries; the theory is comprehensively reviewed in the book by D. Kramer
{\it et al}~\cite{kram}. Here we use notation such that Greek indices are
space-time indices and Latin indices are spatial. We will often use a dot to
denote a time derivative, e.g. $\dot{V}=\partial_t V$.

Let the scalar field $V$ and the null vector field $k_\alpha$ define a
Kerr-Schild metric
\begin{equation}
g_{\alpha \beta}=\eta_{\alpha\beta}-2Vk_{\alpha}k_{\beta},
\label{eq:metg}
\end{equation}
where $k_{\alpha}=(-1,k_i)$ is null. We choose the background to be a
standard inertial coordinate system with
space-time foliated by time slices $t=$ const and the Minkowski metric given
by $\eta_{\alpha \beta}=$diag(-1,1,1,1). The contravariant null vector
satisfies $k^{\alpha}=g^{\alpha\beta}k_{\beta}=\eta^{\alpha\beta}k_{\beta}$.
For the spatial components we write $k^{\alpha}=(1,k^i)$ so that $%
k^i=\delta^{ij}k_j$.
As an example consider the Schwarzschild geometry. In ingoing Kerr-Schild
form it is
\begin{equation}
V=-\frac{M}{r}, \; \; k_i dx^i = -dr.
\end{equation}
Another example is the Kerr solution, which is rather long~\cite{mtw}.

We wish to pose black hole Cauchy data at $t=0$. We only require that
initially the geometry be of the Kerr-Schild type. In particular, we assume
that the evolution of the initial data leads to a metric of the form
$g_{\alpha \beta}+t^2 j_{\alpha \beta}$, with $j_{ab}$ well behaved at $t=0$
and $g_{\alpha \beta}$ given by equation (\ref{eq:metg}).
The initial data at $t=0$ must satisfy the constraint equations
$G_{ab}n^b=0$,
where $n^a$ is the unit normal to the foliation. These
equations do not
contain second time derivatives so that $j_{ab}$ does not enter and we can
analyze them assuming a pure Kerr-Schild form, i.e. setting $j_{ab}=0$. In a
$3+1$ decomposition, these constraints become spatial equations for the
initial values of $V$, $k_i$ and their first time derivative, which
determine the initial values of the 3-geometry and extrinsic curvature. For
the family of global vacuum Kerr-Schild solutions, the null vector field $%
k_{\alpha}$ is necessarily geodesic and shear-free. The shear-free property
makes it unlikely that a globally Kerr-Schild space-time can contain more
than one black hole. Here, for our instantaneously Kerr-Schild spacetimes,
we impose no differential conditions on $k_{\alpha}$ except those that
follow from the null condition, $\partial_{\alpha}( k^i k_i)=0$.

\section{The constraint equations}
\label{sec:cons}

For completeness, we present the equations in covariant form, and also in
``3 + 1'' formalism with partial derivatives.

\subsection{The covariant constraint equations}

Space-time indices take on the values $\alpha,\beta,\gamma, \delta, \ldots =
0,1,2,3.$ The symmetric part of a tensor is $T_{(\alpha \beta)} \equiv \; {%
\frac{1 }{2}} (T_{\alpha \beta}+T_{\beta \alpha})$, and the antisymmetric
part is $T_{[\alpha \beta]} \equiv \; {\frac{1 }{2}} (T_{\alpha
\beta}-T_{\beta \alpha}) .$ The operator for the derivative along the null
congruence $k^\alpha$ is $D \equiv \; _{;\alpha}k^{\alpha}.$
The acceleration vector of the null congruence $k^\alpha$
is $  a^{\alpha} = k^{\alpha}_{\;
;\beta}k^{\beta} = k^{\alpha}_{\; ,\beta}k^{\beta} =Dk^{\alpha}. \nonumber
$ \noindent The square of its norm is $a^2 \equiv a^{\alpha} a_{\alpha}$,
while the twist of the congruence, $\omega,$ is defined by ${\omega}^2
\equiv \; {\frac{1 }{2}} k_{[\alpha; \beta]} k^{\alpha; \beta}$.

We may compute the Ricci tensor in Euclidean coordinates explicitly in terms
of ordinary derivatives as
\begin{eqnarray}
R_{\beta \delta} & = & L_{\beta \delta} + 2V [ (DV) k^{\alpha}_{\; , \alpha}
-2V_{, \alpha} a^{\alpha} + D^{2}V +8V^2 \omega^2] k_{\beta} k_{\delta}
 \nonumber \\
& & +4V[2(DV) + V k^{\alpha}_{\; , \alpha}] a_{(\beta} k_{\delta)}
+2V^{2}a_{\beta} a_{\delta}  \nonumber \\
& & +4V^{2}k_{(\beta}k_{\delta),\alpha \gamma} k^{\alpha}k^{\gamma},
\end{eqnarray}
where
\begin{eqnarray}
L_{\beta \delta} & = & {\frac{1 }{2}} \eta^{\alpha \gamma} [g_{\alpha \delta
, \beta} + g_{\alpha \beta , \delta} - g_{\beta \delta , \alpha}]_{,\gamma}
\nonumber \\
& = & \eta^{\alpha \gamma} (V k_{\beta} k_{\delta})_{,\alpha \gamma} -(V
k^{\gamma} k_{\beta})_{, \delta \gamma} -(V k^{\gamma} k_{\delta})_{, \beta
\gamma}
\end{eqnarray}
and $ D^{2}V = V_{,\alpha \beta} k^{\alpha} k^{\beta} + V_{,\alpha}
a^{\alpha}$. The Ricci scalar is
\begin{equation}
R = -2(V k^{\alpha} k^{\beta})_{,\alpha \beta} + 2 V^{2} a^{2} .
\end{equation}
The Ricci tensor may also be put in covariant form
\begin{eqnarray}
R_{\beta \delta} & = & (V k_{\beta} k_{\delta})^{;\alpha}_{\; \; ;\alpha} -
(V k^{\alpha} k_{\delta})_{;\alpha \beta} - (V k^{\alpha}
k_{\beta})_{;\alpha \delta}  \nonumber \\
& & + 2V [D^{2}V + (V a^{\alpha})_{;\alpha}-a^{2}V^{2}] k_{\beta}
k_{\delta}+ [2V(DV)-V^{2} k^{\alpha}_{\;
\;;\alpha}]D(k_{\beta}k_{\delta}) \nonumber \\
& & +
V^{2}D^{2}(k_{\beta}k_{\delta})+2V^{2}a^{\alpha}k_{\alpha;(\beta}k_{\delta)}
\nonumber
\end{eqnarray}
and the covariant Ricci scalar is
\begin{equation}
R = -2(V k^{\alpha} k^{\beta})_{;\alpha \beta} + 2 V^{2} a^{2}.
\end{equation}
Using the expressions for the Ricci tensor and Ricci scalar above we may
easily write out the Einstein tensor $G_{\beta \delta} = R_{\beta
\delta} - {1 \over 2} R g_{\beta \delta}$. The formulae in this section were
calculated by hand and checked using the computer algebra system MathTensor.

\subsection{The constraint equations in ``3 + 1'' form}

\label{sec:3+1}

It is straightforward to show that the 3-metric $\gamma_{ij}$, the lapse $%
\alpha$ and the shift $\beta_i$ take the following forms:
\begin{equation}
\gamma_{ij}=\delta_{ij}-2 V k_i k_j , \; \; \alpha=\frac{1}{\sqrt{1-2V}}, \;
\; \beta_i=2 V k_i.
\end{equation}
The ADM constraint equations are found from the extrinsic curvature $K_{ij}$%
, which is defined by
\begin{equation}
K_{ij}=\frac{1}{2 \alpha} (\beta_{i|j} + \beta_{j|i} -
{\dot {\gamma}}_{ij}),
\end{equation}
where ${}_|$ denotes covariant differentiation with respect to the metric $%
\gamma_{ij}$. The standard form of the ADM constraints is
\begin{equation}
{}^{(3)}R + (K^i{}_i)^2 - K^i{}_j K^j{}_i = 2 \alpha^2 G^{00} =0, \; \;
K^j{}_{i|j} - (K^j{}_j)_{,i}=\alpha G^0{}_j =0,
\label{eq:adm}
\end{equation}
where the first equation is known as the Hamiltonian constraint, and the
second equation is the momentum constraint; $G^{00}$ and $G^0{}_j$ are
components of the 4-dimensional Einstein tensor. In equation (\ref{eq:adm})
indices are raised by the metric $\gamma^{ij}$, but we will not be using this
convention below.

We have carried out an algebraic computation using REDUCE to evaluate
the constraints for the instantaneous Kerr--Schild metric. Before we
give the results, which are obtained after some simplification of
the REDUCE output, some general comments are in order. The calculation is
carried out in the Euclidean background metric where the components of
covariant and contravariant vectors are identical. Thus it is possible to
write all indices as suffices, and we choose to do this because it greatly
simplifies the computer algebra.
A repeated latin suffix means summation with respect to the metric
$\delta^{ij}$. These conventions will also apply throughout the rest of
this section, and in section~\ref{sec:orth}.  Some simplification rules were
used in the computer algebra, explicitly $k_i k_i =1$, $k_i k_{i,j}=0$ and $%
k_i {\dot {k}}_i=0$. The quantity $z_i$ is defined to be a vector
orthogonal to $k_i$, i.e. $z_i k_i=0$; thus $z_i$ is not uniquely defined
but instead lies in a 2-D surface. The analogue of the 4-acceleration
$a^\alpha$
is denoted by $A_j$ defined by $A_j = k_{j,i} k_i$.

The momentum constraint quantity $\alpha G^0{}_j k_j$ is denoted by
$m_k$ , and is
\begin{eqnarray}
&0& = m_k =\nonumber \\
& &{-\sqrt{-2 V + 1} \over (2 V - 1)} \left[\phantom{1 \over 1}k_{i,jj}
k_i V - \dot{k_i} A_i V\right.
- 2 A_{j,j} V^2 - k_{j,ji} k_i V
- 2 \dot{k}_{j,j} V^2 + \dot{k}_{j,j} V \nonumber \\
& &- V_{,ij} k_i k_j + V_{,jj}
+ \dot{V} k_{j,j} - (4 V + 1) V_{,j} A_j
+ 2 A_j A_j V^2 - k_{j,j}V_{,i} k_i\nonumber \\
& &\left.- \dot{k}_j ((4 V -1) V_{,j} - 2 A_j
V^2)\phantom{1 \over 1} \right].
\label{eq:mk}
\end{eqnarray}
In order to eliminate $\dot{V}$, the Hamiltonian constraint is not
expressed in the standard ADM form. Rather, we use $G^0{}_0$, which is
\begin{equation}
G^0{}_0=-\alpha^2 G^{00} + \alpha 2 V m_k.
\end{equation}
Denoting $G^0{}_0$ by $h_2$, we find
\begin{eqnarray}
&0& = h_2 = \nonumber \\
& &{-1 \over (2 (2 V - 1)^2)}
{\Large \left[\phantom{1\over
1}\right.}\left(\right. V_{,i}
A_i + 2 k_{i,j} k_{j,i} V - 4 k_{j,ji} k_i V^2
- 8A_{j,j} V^3\nonumber \\ & &+ 4 k_{i,ji} k_j V
- 2 k_{j,ii} k_j V - 4 \dot{k}_i A_i V^2
+ 4 k_{i,jj} k_i V^2\left.\right)
   (2 V - 1)\nonumber \\
& &-\left\{\phantom{1\over 1} 2 \left[\phantom{1\over 1}
   (k_{i,i} V + 2 V_{,i} k_i) k_{j,j}\right.
    \right.
- (k_{i,j}k_{i,j} V + 1) - V_{,jj} + V_{,ij} k_i k_j\nonumber \\
& &\left. + 2 \dot{k}_{j,j} V^2 + \dot{k}_j \dot{k}_j
V^2\phantom{1\over
   1}\right] (2 V -
   1)^2\nonumber \\ & &+ 2 (2 V - 1)^2
- V_{,i}V_{,j} k_i k_j + V_{,j} V_{,j} - 4 A_{j,i} k_j k_i
   V^2
\left. + 4 k_{i,jl} k_i k_j k_l V^2\phantom{1\over 1}\right\}\nonumber \\
& &+ 2 (4 V - 1) (V -1) A_j{}A_j V^2
- (32 V^3 - 16 V^2 - 8 V +3) V_{,j} A_j\nonumber \\
& &- 4 (4 V_{,j} V - 2 V_{,j} - A_j V) (2 V - 1)
   \dot{k}_j V {\Large \left. \phantom{1 \over 1}\right]}.
\label{eq:h2}
 \end{eqnarray}
The momentum constraint quantity $\alpha G^0{}_j z_j$ is denoted by
$m_z$ , and is
\begin{eqnarray}
&0& = m_z =\nonumber \\
& &{\sqrt{(- 2 V + 1)} \over (2V - 1)^2}
 {\Large \left[\phantom{1 \over
   1}\right.}\left(\phantom{1
   \over 1}V_{,i}k_{j,j} - \dot{V}A_i +
   \dot{V}_{,i}\right.\nonumber \\
& &\left. + k_{j,ji} V - k_{i,jj} V - \dot{k}_{i,j} k_j V
   \phantom{1 \over 1}\right) (2
   V - 1)\nonumber \\
& &+ 2 (V_{,j}k_j)_{,i} V - 2 (V_{,j}k_j)_{,i} + V_{,ij}
   k_j
- (k_{j,j} V + V_{,j}k_j) (2 V - 1) \dot{k}_i \nonumber \\
& &- [2 (V_{,j} - A_j V^2) - (2 V - 1) \dot{k}_j V] (2 V - 1)
   k_{i,j}\nonumber \\
& &- [2 (\dot{k}_j + A_j) (2 V - 1) V^2 - V_{,j}]
   k_{j,i}{\Large \left. \phantom{1 \over 1}\right]} z_i.
\label{eq:mz}
\end{eqnarray}

\section{On the solution of the constraint equations}
\label{sec:sol}

\subsection{The surface orthogonal case}
\label{sec:orth}

In the case of a black hole with spin, such as a Kerr black hole, $k_i$ is
proportional to $\nabla_i \phi + c_i$ with $c_i$ tangent to the surfaces
given by $\phi =$ const. While the analysis of the solution of the
constraint
equations in this general case is quite complicated it
simplifies drastically if $c_i=0$ so that $k_i$ is orthogonal to 2-surfaces $%
S$, with
\begin{equation}
k_i=\nabla_i \phi / |\nabla \phi|.
\label{eq:kphi}
\end{equation}
The constraints may then be
solved by carrying out a hierarchy of integrals along the rays normal to the
surfaces $S$. (Note that difficulties may arise at points where $\nabla_i
\phi=0$). Here, we confine our attention to this surface orthogonal
case. It includes the description of boosted black holes.

Formally, the ansatz used is that $\phi$ is given everywhere on a 3-D
manifold with Euclidean metric $\delta_{ij}$. The vector field $k_i$ is
defined by equation (\ref{eq:kphi}), and is everywhere
orthogonal to 2-surfaces $S$ (which are equipotentials of $\phi$).
Also, $V$ and $\dot{k}_i$ are given on a particular equipotential $S_0$.
The solution procedure assumes that $k_i$ is defined everywhere, and also that
$k_{i,i} \neq 0$ everywhere. In multiple black hole problems it is likely
that these conditions could be violated at isolated points, and special
procedures would need to be developed for integrating through such points.

The equations $h_2=0$, $m_k=0$ and $m_z=0$ appear complicated, but it should be
remembered that we know terms involving $k_i$ and any spatial derivative,
e.g. $A_{i,i}$. In addition, we know
terms involving $V$, $\dot{k}_i$ and any spatial derivative within $S_0$. In
particular, the following terms are known on $S_0$  from the given data:
\begin{itemize}
\item $V_{,i} A_i$ and
$V_{,j} \dot{k}_j$. Since both $A_i$ and $\dot{k}_j$ are
orthogonal to $k_i$ they both lie in $S_0$, and therefore the $V$-derivative
in both terms is calculable from data on $S_0$.
\item $V_{,jj}-V_{,ij}k_ik_j$
and $-k_jV_{,j}k_iV_{,i}+(V_{,j})^2$.
In both cases the off-$S_0$ derivatives cancel out, and so the terms are
calculable from data on $S_0$.
\item $\dot{k}_{i,i}$. Since $\dot{k}_i$ is orthogonal to
$k_i$, it follows that $\dot{k}_{i,i}$ can be expressed in terms of
derivatives of $\dot{k}_i$ within $S_0$.
\end{itemize}
Equation (\ref{eq:h2}) ($h_2=0$) is a linear equation for $k_i V_{,i}$, and may
be written schematically as
\begin{equation}
\partial_{\mbox{\boldmath$k$}} V = {\cal{O}}_1 (D_1),
\end{equation}
where ${\cal{O}}$ represents an operator on $S_0$, and $D$ represents the data
given on $S_0$; $\partial_{\mbox{\boldmath$k$}}$ means $k_i \partial_i$. Next,
$\dot V$ is found explicitly from equation (\ref{eq:mk})
($m_k=0$):
\begin{equation}
 \dot V = {\cal{O}}_2 (D_2,\partial_{\mbox{\boldmath$k$}} V ).
\end{equation}
Finally, equation (\ref{eq:mz}) ($m_z=0$) is regarded as a linear equation to
determine $\dot{k}_{i,j}k_jz_i$ (Noting that $z_i \partial_i$ means a
derivative
within $S_0$, it is straightforward to see that all terms are expressible
entirely in terms of known data):
\begin{equation}
z_i  \partial_{\mbox{\boldmath$k$}} \dot{k_i} =
{\cal{O}}_3 (D_3, \partial_{\mbox{\boldmath$k$}} V, \dot V ).
\label{eq:mz1}
\end{equation}
Because $z_i$ can take two independent directions, equation (\ref{eq:mz1})
gives
two components of $\dot{k}_{i,j} k_j$. The remaining component follows from
the condition that $\dot{k}_i k_i=0$ so that $\dot{k}_{i,j} k_i = - k_{i,j}
\dot{k}_i$, and therefore $\dot{k}_{i,j} k_j k_i = - k_{i,j} k_j \dot{k}_i$.

In this way $\dot V$ and the $\partial_{\mbox{\boldmath$k$}}$ derivatives
of $V$ and $\dot{k}_j$ are found at each point of
$S_0$. Thus the solution may
be extended to the ``next'' 2-surface S, and, in the absence of
singularities, to the whole manifold.

\subsection{Multiple black holes}

\label{sec:mult}

For a single (unboosted) Schwarzschild black hole, $k_i$ is a radial field
which can be obtained by the equipotentials of the function $\phi =1/r$.
Thus the choice $\phi = \Sigma_i M_i/r_i$ generates Schwarzschild-like
equipotentials about each site where $r_i =0$ and becomes a candidate for
data for (unboosted) multiple black holes with masses $M_i$. By giving
appropriate data for $V$ on $S_0$ we shall show that this does produce the
necessary apparent horizons.

In order to describe boosted black holes we begin with the null vector
$k_{\alpha} =-\nabla_{\alpha}(t+r)$ for a single Schwarzschild black hole.
Under a boost with velocity $v$
in the $z$-direction, $t\rightarrow (t+vz)/\sqrt{1-v^2}$ and
$z\rightarrow (z+vt)/\sqrt{1-v^2}$, we have
\begin{equation}
    t+r\rightarrow {t+vz+\sqrt{(1-v^2)(x^2+y^2)+(z+vt)^2}
              \over \sqrt{1-v^2}}.
\end{equation}
Then, for a single boosted Schwarzschild black hole, initial data for
$k_i$ at $t=0$ can be obtained from the equipotentials of
${\hat \phi}=\{vz+[(1-v^2)(x^2+y^2)+z^2]^{1/2}\}^{-1}$

Construction of a solution corresponding to multiple black holes
requires a numerical integration procedure. For
simplicity, we restrict our discussion to an analysis of the binary case
$\phi = M_1/r_1 +M_2/r_2$, although the above considerations make it apparent
how to generalize to the case of multiple, boosted black holes. In this
simple case, the equipotentials consist of small disjoint (topological)
spheres which merge in a ``figure-eight'' to form a common set of outer
spheres. The hard problems arise at the bifurcation point of the
``figure-eight''. We could avoid this problem by considering the case where
the two holes are close so that they are surrounded by a common apparent
horizon. This would suffice to determine data for the exterior spacetime
without going down to the bifurcation point. The problem with doing this is
that we would probably not be able to include the merger radiation this way
and that is the most interesting feature. Nevertheless, it seems advisable to
tackle the close binary case first.

\subsection{Apparent horizons}

In order to make sure that the data proposed above really represents black
hole data we must demonstrate the existence of apparent horizons. Consider
the equipotential 2-surfaces $S$ given by $\phi =$ const. For the
reasons
given in the introduction, we are interested in ingoing Kerr-Schild
coordinates so that $k^{\alpha}$ is chosen to be the incoming null vector
normal to $S$. The divergence of $k^{\alpha}$ is the same when measured
either in the full Kerr-Schild metric or in the background Minkowski metric.
Thus it is apparent from the asymptotically spherical shape of the
equipotentials on the approach to each $r_i=0$ site that $\rho_{\mbox{in}}$
(the
incoming divergence of $S$) goes uniformly negative.

The outgoing null normal to $S$ is given by $\ell^{\alpha}=2t^{%
\alpha}-(1+2V)k^{\alpha}$, where $t^{\alpha}=(1,0,0,0)$ is the time
translation Killing vector of the Minkowski background. It has normalization
$\ell^{\alpha}k_{\alpha}=-2$. Let
$(\ell^{\alpha},k^{\alpha},m^{\alpha},{\bar m}^{\alpha})$
form a null tetrad. We choose normalization $m^{\alpha}{\bar m}_{\alpha}=2$
(with respect to both the Kerr-Schild and Minkowski metrics).The complex
spatial vector $m^{\alpha}$ is tangent to $S$. The outward divergence of $S$
is given by $\rho_{\mbox{out}}=m^{(\alpha}{\bar
m}^{\beta)}\nabla_\alpha\ell_{\beta}
$. We have
\begin{equation}
2m^{(\alpha}{\bar m}^{\beta)}\nabla_\alpha t_{\beta}= m^{(\alpha}{\bar m}%
^{\beta)}{\cal L}_t g_{\alpha\beta}= m^{(\alpha}{\bar m}^{\beta)}{\cal L}_t
\eta_{\alpha\beta}=0,
\end{equation}
since $t^{\alpha}$ is a Minkowski Killing vector. As a result, $%
\rho_{\mbox{out}}=-(1+2V)\rho_{\mbox{in}}$.

Where $|V|$ is small, $\rho_{\mbox{in}}$ is negative and $\rho_{\mbox{out}}$ is
positive,
as for a sphere in Minkowski space. For the Schwarzschild solution $V=-M/r$
and the (apparent) horizon forms at $V=-1/2$. Our result shows this holds in
general: An equipotential with $V=-1/2$ is marginally trapped and determines
the position of the apparent horizon on the initial time slice.

Thus, since $V$ is free data on some choice of equipotential $S_0$ (which
might consist of disjoint pieces), it is straightforward to give data that
guarantees the existence of black holes. Note that the location of apparent
horizons is considerably simpler here than in the standard approach. This
suggests that the apparent horizon itself be chosen as the inner
equipotential to start the integration process.

\section{Perturbative calculation}
\label{sec:pert}

We use the above methods to find initial data for a {\it surface orthogonal}
problem, specifically a perturbed Schwarzschild geometry.

\subsection{The close approximation}

As described in Sec. \ref{sec:mult}, two non-spinning black holes initially
at rest can be modeled in terms of the potential $\phi = M_1/r_1 +M_2/r_2$,
where $r_1=|x^i-x_1^i|$ and $r_2=|x^i-x_2^i|$. Without loss of generality,
we can take the line connecting the black holes to be the $z$-axis, with the
origin located according to the center of mass condition $M_1z_1+M_2z_2 =0$.
We let  $a=z_2-z_1$ and $M=M_1+M_2$, and use standard spherical
polar coordinates $(r,\theta,\varphi)$. The close approximation is defined by
the condition
\begin{equation}
\epsilon = \frac{a}{M} \ll 1.
\label{eq:eps}
\end{equation}
We find the expansion
\begin{equation}
\phi ={\frac{M}{r}} +{\frac{\epsilon^2 M_1 M_2 M P_2}{r^3}}
\end{equation}
with $O(\epsilon^3)$ remainder, where
\begin{equation}
P_2=(3 \cos^2\theta-1)/2.
\end{equation}
The approximation is valid when $r>$ max($z_1,z_2$), and improves with
increasing $r$. To this order of approximation,
the  covariant components of ${\mbox{\boldmath $k$}}$ in standard spherical
coordinates are
\begin{equation}
(k_r,k_\theta,k_\varphi)=(-1, -{\frac{3\epsilon^2 M_1 M_2 \cos\theta
\sin\theta}{r}},0).
\label{eq:kr}
\end{equation}
The ansatz on the initial hypersurface $t=0$ is then equation (\ref{eq:kr})
above,
together with
\begin{equation}
V=-M/r+\epsilon^2 v(r) P_2
\end{equation}
\begin{equation}
\dot V=\epsilon^2 v_T(r) P_2
\end{equation}
\begin{equation}
(\dot k_r,\dot k_\theta,\dot k_\varphi)=
          (0,3 \epsilon^2 k_T(r) \cos\theta \sin\theta,0).
\end{equation}
The functions $v_T(r)$ and $k_T(r)$ are, in a 4-dimensional sense, time
derivatives; but since
the whole analysis is carried out in a spacelike 3-D manifold, formally we
regard
$v_T(r)$ and $k_T(r)$ as independent functions on the manifold rather than as
derivatives. The objective now is to determine the functions $v(r)$, $v_T(r)$
and
$k_T(r)$.

\subsection{The algebraic computations}

The computation of the constraint equations has been done in two different
ways, and using two different algebra systems (Maple and REDUCE). In one
calculation we put the ansatz above into the equations reported in section
\ref{sec:3+1}, and in the other calculation the ansatz is put into the full
4-dimensional Einstein equations and $G^0{}_\alpha$ is found. Both
procedures
give identical results, and this serves as a partial confirmation that our
computer algebra programs are correct.

The constraint equations $m_k=0$, $h_2=0$ and $m_z=0$ were evaluated, and to
order zero in $\epsilon$ they were satisfied identically; and to order
$\epsilon^2$ the angular terms cancelled out and the equations involved $r$
only. After some manipulation we find:
\begin{equation}
v_T(r)={-1 \over r^3} \left[ 6 k_T(r) M^{2} + 3 M r
k_T(r)
+ \frac{6 M_1 M_2 M}{r^2} \{ r-M \} + 3 v(r) r^{2} \right],
\label{eq:vT}
\end{equation}
\begin{equation}
v'(r) ={-1 \over r^3} \left[ 4 v(r) r^{2}
+ \frac{6 M_1 M_2 M}{r^2} \{ r-M \}+ 6
k_T(r) M^{2} \right], \label{eq:dv}
\end{equation}
\begin{equation}
k_T'(r) ={1 \over Mr^2} \left[ 2 M r k_T (r)
-\frac{6 M_1 M_2 M}{r}- v(r) r^{2} \right],
\label{eq:kT}
\end{equation}
where $'$ represents ${d \over dr}$.
Equations (\ref{eq:dv}) and (\ref{eq:kT}) constitute a system of two
first-order equations for $v'(r)$ and $k_T'(r)$; once they have been solved
$v_T(r)$ is found explicitly from equation (\ref{eq:vT}). The system
(\ref{eq:dv}) and (\ref{eq:kT}) may be re-expressed as one second-order
equation:
\begin{equation}
v''(r) r^{5} + 5 r^{4}
v'(r)  - 6 r^{2} M v(r) -\frac{6 M_1 M_2 M}{r} \{3r +2M\}=0.
\label{eq:d2v} \end{equation}

\subsection{Behavior of the solution}

In this section we are concerned with finding the behavior of the solution
of equation (\ref{eq:d2v}).
Making the transformation
\begin{equation}
r=\frac{24 M}{x^2}, \; \; v(r)=x^4 u(x),
\label{eq:tr}
\end{equation}
equation (\ref{eq:d2v}) becomes
\begin{equation}
x^2 u_{,xx} + x u_{,x} -u (x^2 + 16) - \frac{M_1 M_2}{6912 M^2} (36 x^2
+x^4)=0.
\label{eq:d2u}
\end{equation}
The solution to equation (\ref{eq:d2u}) is
\begin{equation}
u(x)=C_1 I_4(x) + C_2 Y_4(ix) + \frac{M_1 M_2}{M^2} (w(x) - \frac{x^2}{2304}
-\frac{I_4(x) \log x}{72} ),
\label{eq:usol}
\end{equation}
where $I_4$ and $Y_4$ are standard Bessel functions~\cite{num}
and $i=\sqrt{-1}$. The function $w(x)$ may be written as
\begin{equation}
w(x)=w_6 x^6 + w_8 x^8 + w_{10} x^{10} + ...
\end{equation}
and the coefficients in the series are found from the following recurrence
relation applied to $n=$6, 8, 10..., and using the initial condition $w_4=0$:
\begin{equation}
(n^2-16) w_n = w_{n-2} + \frac{n}{36 (\frac{n}{2}-2)! (\frac{n}{2}+2)! 2^n}.
\end{equation}
Using equation (\ref{eq:tr}) we may now find $v(r)$; then
equations (\ref{eq:dv}) and (\ref{eq:vT}) lead to $k_T(r)$ and $v_T(r)$.

The leading term in the inhomogeneous part of the solution (\ref{eq:usol})
behaves as $x^2$ as $x \rightarrow 0$,
which means that $v(r)$ would behave as $r^{-3}$ as $r \rightarrow \infty$;
this is the expected quadrupole behavior at infinity.
The constants $C_1$ and $C_2$ represent gauge freedom, as is evident from the
fact that they persist in the Schwarzschild case ($M_1=M$ and $M_2=0$).
$Y_4$ behaves as $x^{-4}$ near $x=0$, so that if $C_2 \neq 0$ then
$v(r)$ would behave as a constant as $r \rightarrow \infty$. While this is
possible, equation (\ref{eq:dv}) then indicates that $k_T(r)$ would be
singular as $r \rightarrow \infty$. Thus it is convenient to fix $C_2=0$.
$I_4$ behaves as $x^4$ near $x=0$, so that $v(r)$ would behave as $r^{-4}$ as
$r \rightarrow \infty$. $C_1$ could be chosen so as to fix the position of
the apparent horizon: for example, $C_1$ could be chosen so that $v(2M)=0$,
which would mean that the position of the apparent horizon is not perturbed
relative to the Schwarzschild case.

\section{Conclusion}
\label{sec:conc}

In this paper we have shown how to express the initial value problem of
general relativity in a way that can be solved using initial data
that is instantaneously Kerr-Schild. We have used this approach
to formulate, and solve analytically, the initial value problem of a perturbed
Schwarzschild geometry. It may be interesting to see if the difference
between this initial data and that obtained from the conformally flat method
affects results about the perturbative evolution of a black hole~\cite{pp}.
We have shown that the apparent horizon is described in a very simple way,
$V=-\frac{1}{2}$; thus the generation of initial data for black holes by the
Kerr-Schild approach would automatically give the initial location of the
apparent horizon.

Clearly the Kerr-Schild approach to the initial value problem is very much
in an early stage of development, particularly in comparison with the
conformally flat method that was introduced in 1963~\cite{mis}. This paper
is just a first step towards developing this new approach. The next steps,
which are currently under investigation, will be the construction of a code
to solve numerically equations (\ref{eq:mk}), (\ref{eq:h2}) and (\ref{eq:mz}),
and the generalization of our approach to be able to include problems with
spin, where $k_i$ is not orthogonal to the 2-surfaces $S$.

\acknowledgements

This work has been supported by the Binary Black Hole Grand Challenge
Alliance, NSF PHY/ASC 9318152 (ARPA supplemented) and by NSF PHY
9510895 and NSF INT 9515257 to the University of Pittsburgh. N.T.B. and
M.M. thank the Foundation for Research Development, South Africa, for
financial support. N.T.B., M.M., and R.A.I. thank the University of
Pittsburgh for hospitality. J.W. and R.A.I. thank the Universities of
South Africa and of Durban-Westville for their hospitality.

\end{document}